\documentclass[aps,prb, twocolumn, superscriptaddress,showpacs]{revtex4-1}
\usepackage{graphicx}
\usepackage{amsmath}
\usepackage{amsfonts}
\usepackage{amsthm}
\usepackage{amssymb}
\usepackage{amsbsy}
\usepackage{wasysym}
\usepackage{bm}
\usepackage{mathrsfs}
\usepackage{color}
\usepackage{hyperref}
\usepackage{chngcntr}

\newcommand{\beq}{\begin{equation}}
\newcommand{\eeq}{\end{equation}}
\newcommand{\bea}{\begin{eqnarray}}
\newcommand{\eea}{\end{eqnarray}}

\newlength{\myL}

\def\ket#1{{\left|#1\right\rangle}}
\def\bra#1{{\left\langle #1 \right|}}

\def\be{\begin{eqnarray}}
\def\ee{\end{eqnarray}}

\begin{document}

\title{Quantum revivals and many-body localization}
\author {R. Vasseur}
\affiliation{Department of Physics, University of California, Berkeley, CA 94720, USA}
\affiliation{Materials Science Division, Lawrence Berkeley National Laboratories, Berkeley, CA 94720}
\author{S. A. Parameswaran}
\affiliation{Department of Physics and Astronomy, University of California, Irvine, CA 92697, USA}
\author{J. E. Moore}
\affiliation{Department of Physics, University of California, Berkeley, CA 94720, USA}
\affiliation{Materials Science Division, Lawrence Berkeley National Laboratories, Berkeley, CA 94720}

\date{\today}
\begin{abstract} 

We show that the magnetization of a single `qubit' spin weakly coupled to an otherwise isolated disordered spin chain 
 exhibits periodic revivals in the localized regime, and retains an imprint of its initial magnetization at infinite time. We demonstrate that the revival rate is strongly suppressed upon adding interactions after a time scale corresponding to the onset of the dephasing that distinguishes many-body localized phases from Anderson insulators.  In contrast, the ergodic phase acts as a bath for the qubit, with no revivals visible on the time scales studied.  The suppression of quantum revivals of local observables provides a quantitative, experimentally observable alternative to entanglement growth as a measure of the `non-ergodic but dephasing' nature of many-body localized systems.

\end{abstract}


\maketitle

 The study of the dynamics of closed quantum many-body systems has enjoyed a recent renaissance, driven in part by the ability to cool and trap large collections of atoms and molecules, 
 tune inter-particle interactions, and probe the resulting phases and their dynamics with high spatial and temporal resolution~\cite{RevModPhys.80.885}. An implicit assumption often made in applying the standard tools of many-body theory and quantum statistical mechanics to these isolated systems is the {\it eigenstate thermalization hypothesis} (ETH)~\cite{PhysRevE.50.888, PhysRevA.43.2046}.  The idea of the ETH is that while the dynamics of a quantum mechanical eigenstate are of course not ergodic, individual regions will, in the limit that their size remains much smaller than the whole in the thermodynamic limit, display behavior expected of an ergodic system at a temperature proportional to their initial energy density. However, there exists a class of generic, non-integrable systems that fail to thermalize~\cite{PhysRevLett.106.040401, PhysRevE.85.050102, 0295-5075-101-3-37003} in the sense of ETH, and instead exhibit many-body localization (MBL)~\cite{Basko20061126,revRahul}.
 
In seminal work, Basko {\it et. al.}~\cite{Basko20061126} gave strong arguments for the existence of an MBL phase by examining the stability of the Anderson-localized ({\it i.e.}, non-interacting) phase of disordered electrons~\cite{PhysRev.109.1492} against delocalization by interactions. Apart from the fundamental interest in understanding quantum systems for which the notion of equilibrium is intrinsically absent, many-body localization offers the ability to realize phenomena forbidden by thermodynamic arguments, such as long-range order  below the lower critical dimension~\cite{PhysRevB.88.014206,PhysRevLett.113.107204}, and topologically protected quantum coherent behavior~\cite{2013arXiv1307.4092B,PhysRevB.88.014206,BauerNayak,PhysRevB.89.144201} at non-zero energy density. Several other questions, such as whether MBL can arise in translationally invariant systems~\cite{1742-5468-2014-10-P10010,4893505,PhysRevB.90.165137,2014arXiv1405.5780H} or survive in situations where the single-particle spectrum includes a set of extended states~\cite{PhysRevB.90.195115}, remain subjects of active study (see also {\it e.g.}~\cite{Aizenman:2009aa, Wang:2009aa,  2014arXiv1403.7837I} for recent mathematical developments).

A corollary of the breakdown of ETH is that an MBL system cannot act as a thermalizing bath for an otherwise isolated quantum impurity. 
Considering MBL systems as quantum reservoirs defines a set of unusual quantum impurity problems~\cite{PhysRevB.90.224203}, in which the impurity remains weakly perturbed by the reservoir even when the latter is at infinite effective temperature. Therefore, in this paper we analyze the dynamics induced in a single, initially magnetized test qubit when it is coupled to a disordered spin chain. We study the dynamics following this `quench' via numerical studies and analytical arguments, and demonstrate the distinctive nature  of dissipation and dephasing induced in the qubit depending on whether the spin chain is ergodic, Anderson-localized, or MBL.

We focus on revivals (returns of a time-dependent observable `sufficiently close' to its initial value after deviating `sufficiently far' from it, to be made precise below) of the magnetization of a single qubit.  Specifically, we demonstrate that the constant qubit revival rate in the Anderson insulator is changed to a universal logarithmic decay upon adding interactions. This in turn can be precisely related to the dephasing mechanism responsible for the slow, logarithmic growth of entanglement in the MBL phase~\cite{deChiara:2006p1, PhysRevB.77.064426, PhysRevLett.109.017202, PhysRevLett.110.260601,PhysRevB.90.064201}, compared to saturation in the Anderson case.  
Famously, the existence of revivals in a system of finite phase space is required by Poincare's theorem~\cite{PoincareRecurrence} for (classical) Hamiltonian systems; for systems that move ergodically over some subregion of phase space, the rate of revivals depends inversely on the volume of phase space explored.  It is this volume that differs between Anderson-localized, MBL, and ergodic phases.

 Several prior studies have shed some light on the nature of the many-body localized phase: the existence of a sharp MBL transition and its persistence to infinite temperatures~\cite{PhysRevB.75.155111, PhysRevB.82.174411}, area- rather than volume-law entanglement of MBL eigenstates~\cite{PhysRevLett.108.176803,BauerNayak}, and a phenomenology in terms of quasi-localized conserved quantities~\cite{PhysRevLett.111.127201,PhysRevB.90.174202,2013arXiv1307.0507S}.  
 Unfortunately, much of the intuition about MBL comes from measures, such as entanglement~\cite{deChiara:2006p1, PhysRevB.77.064426, PhysRevLett.109.017202, PhysRevLett.110.260601} or two-point measurements~\cite{PhysRevLett.113.147204}, that are challenging to probe experimentally; therefore, improving the understanding of exactly how much information can be extracted from more conventional measurements is crucial.

\begin{figure}
\includegraphics[width=1\columnwidth]{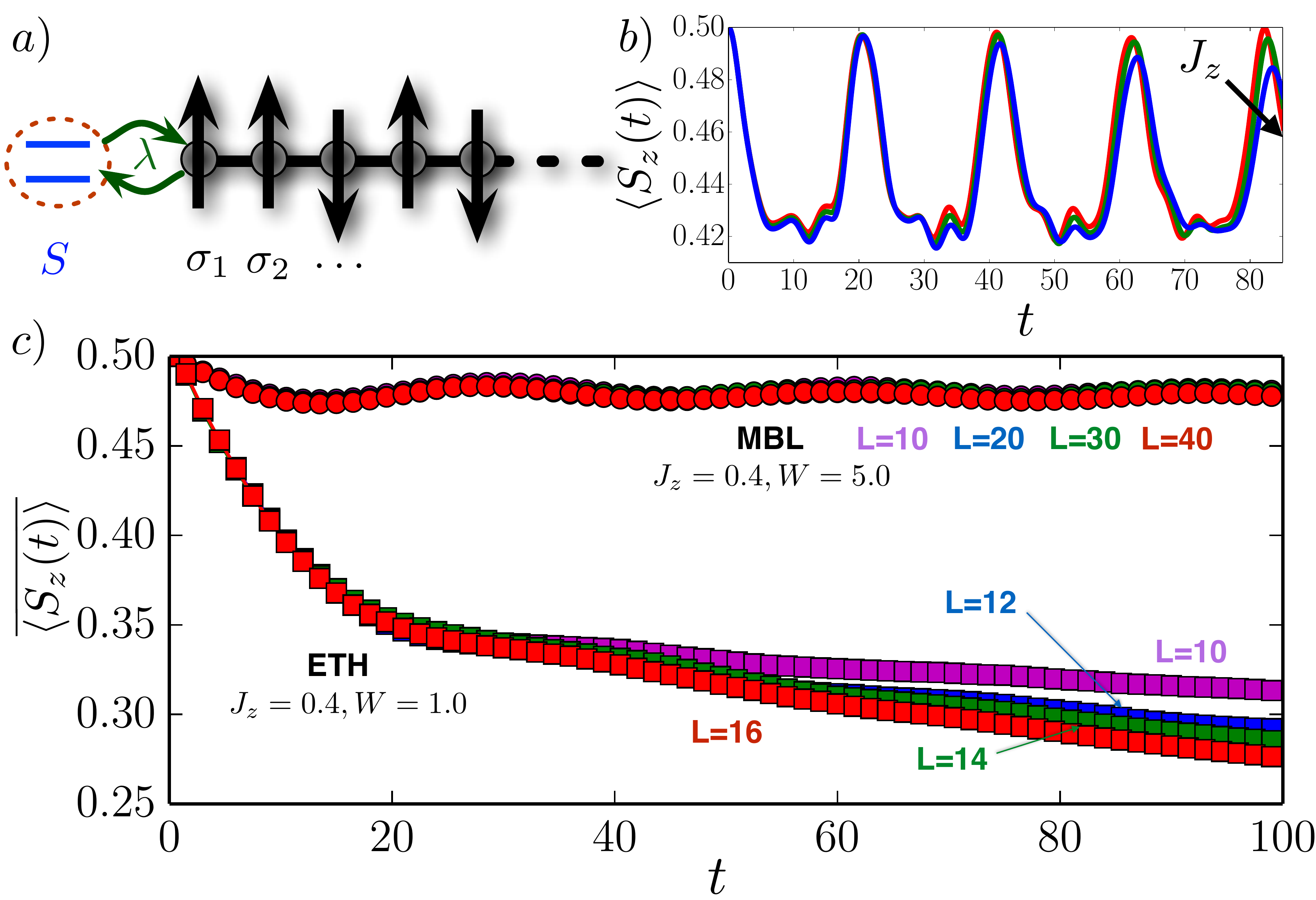}
\caption{\label{fig:Setup} {\bf Quench Protocol and Magnetization Dynamics.} (a) We consider the post-quench dynamics of a two-level `qubit' $\mathbf{S}$ coupled to a one-dimensional chain of atoms with strength $\lambda$ (see Eq.(\ref{eq:ham_xxz})).  (b) Time series for a single instance of disorder, and influence of the interactions strength $J_z=0.0, 0.025$ and $0.05$ on the revivals. (c) Averaged time evolution in the ergodic and MBL phases. Observe that the disorder-averaged magnetization $\overline{\langle S^z(t)\rangle}$ is only slightly diminished with little or no finite-size scaling in the MBL phase, in contrast to the ergodic phase where the magnetization is significantly lower at long times, scaling to zero as $L\rightarrow\infty$. Note that the weak oscillations in the MBL phase scale with $\lambda^{-1}$, and were found to be apparently independent of any localization physics.}
\end{figure}

{\bf The Model.} We will consider a system with a total of $L$ sites, consisting of a single spin-$\frac{1}{2}$ impurity $\vec{S}$ (the `qubit') weakly coupled to an $(L-1)$-site disordered spin-$\frac{1}{2}$ XXZ chain (see Fig.~\ref{fig:Setup}(a)), described by the Hamiltonian
\be\label{eq:ham_xxz}
H &=& H_{\text{XXZ}}[\{\sigma_i\}] + \frac\lambda4 \left(S^+\sigma_1^- + S^-\sigma_1^+\right),
\\ H_{\text{XXZ}} &=& \sum_{i=1}^{L-1}\frac{J_\perp}{8}\left(\sigma_i^+\sigma_{i+1}^- + \sigma_i^-\sigma_{i+1}^+\right) +\frac{J_z }{4}\sigma^z_i \sigma^z_{i+1} +\frac{h_i}{2}\sigma^z_i,\nonumber
\ee
where the $\sigma_i$ are Pauli matrices, and the random fields $h_i$ are drawn randomly from the interval $[-W,W]$ .
Throughout, we will set $J_\perp=1$, and restrict ourselves to even $L$.
This problem is equivalent, via a Jordan-Wigner transformation, to a model of spinless fermions hopping in the presence of onsite disorder  and nearest-neighbor interaction $J_z$, with the end of the chain coupled to a single impurity level. For $J_z =0$ and arbitrarily small $W$, every eigenstate is Anderson localized, as appropriate to a  one-dimensional noninteracting disordered system.  We will study the different phases of (\ref{eq:ham_xxz}) and their corresponding dynamics as the strength of disorder and the interactions are varied. We note that it {\it is} crucial to be able to address the `qubit' and polarize it in the initial state, and subsequently tune its on-site field to zero, in order to study the magnetization dynamics in the fashion probed here. In this sense the qubit is distinct from the rest of the chain; from now on we set $\lambda=0.2$.

We simulate~\cite{SupMat} the time evolution governed by (\ref{eq:ham_xxz}), following a global quench from an initial condition in which the qubit is initialized to be `up', while the `bulk' spins are in initially in a state $\ket{\psi_0}$:
\be
\ket{\Psi(t=0)} \equiv\ket{\Psi_0} = \ket{\uparrow}_{S_z}\otimes \ket{\psi_0[\{\sigma_i\}]}.
\ee
We consider two different alternatives for $\ket{\psi_0[\{\sigma_i\}]}$: either a completely random  $\sigma_i^z$ product state, or else a random product state constrained to have total magnetization $S^z_\text{tot} =0$; while the latter choice results in significantly smaller error bars, the results are otherwise independent of this choice, and indeed we expect that any initial state of sufficiently high energy density should yield similar results to those reported here.

\begin{figure}
\includegraphics[width=\columnwidth]{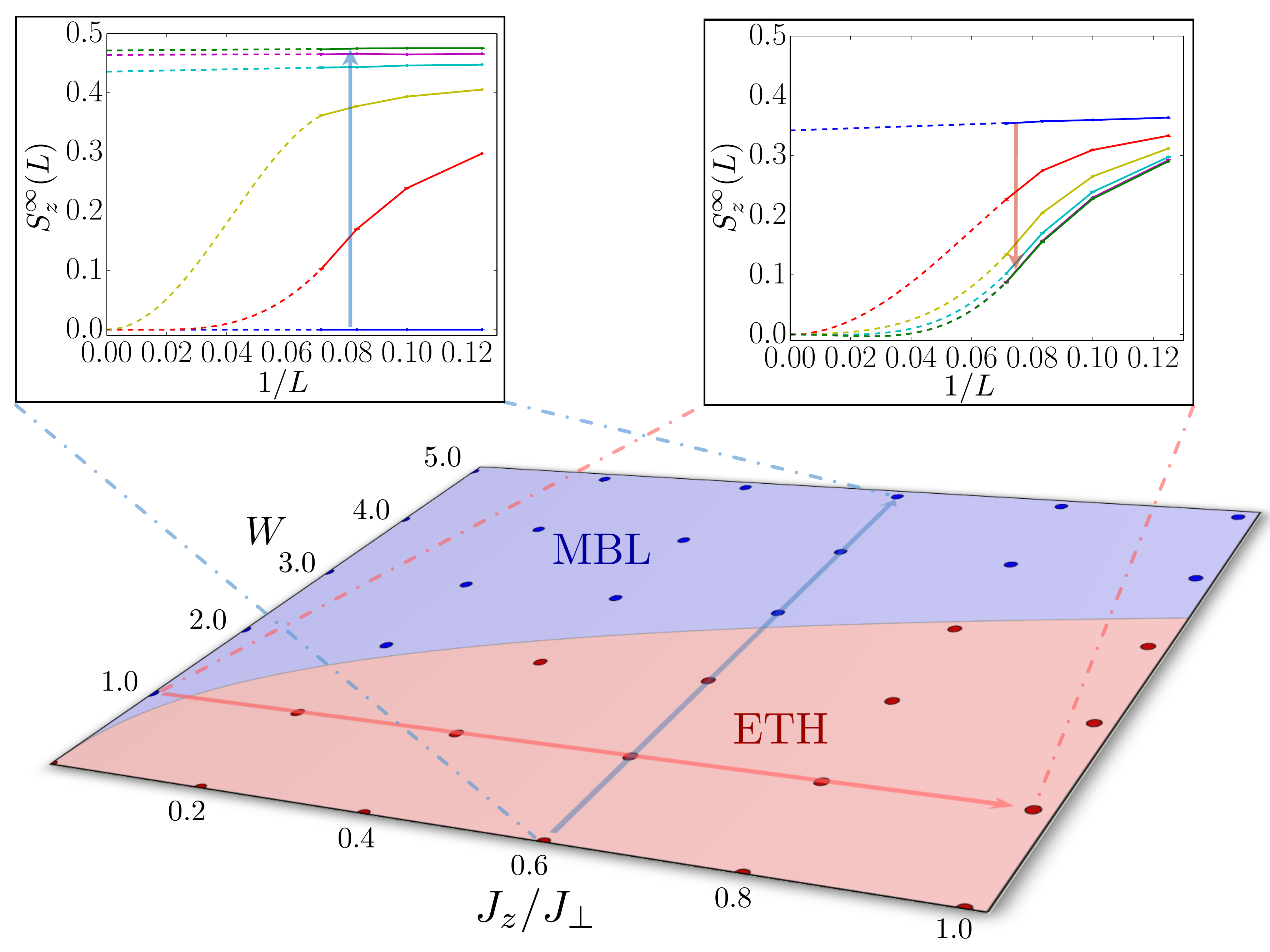}
\caption{\label{fig:phases}{\bf Phase diagram of the disordered XXZ chain}, obtained by finite-size-scaling analysis of infinite-time qubit magnetization $\overline{S^z_\infty}$. (Inset) Sample finite-size scaling for two representative cuts, shown. The dashed extrapolations are shown to guide the eye, see~\cite{SupMat} for a detailed finite size analysis. }
\end{figure}

{\bf Phase Diagram.} As a first step, we establish the phase diagram of the disordered XXZ chain by examining the long-time behavior of the qubit magnetization (see also Ref.~\onlinecite{PhysRevB.89.220201}). To do so, we use the numerically computed exact eigenstates $\ket{\alpha}$ of (\ref{eq:ham_xxz}) 
for a given disorder realization and then average over disorder to obtain
\be\label{eq:InfM}
\overline{S^z_\infty} \equiv \overline{\sum_{\alpha} \bra{\alpha}S^z\ket{\alpha}  \left|\langle \Psi_0 | \alpha\rangle\right|^2}, 
\ee
where $\ket{\Psi_0}$ is one of the choices of initial state above, and the bar denotes disorder-averaging. In the MBL phase, finite-size extrapolation of $\overline{S^z_\infty}$ to $L\rightarrow\infty$ yields a nonzero constant for the infinite-time qubit magnetization. In contrast, in the ergodic phase $\overline{S^z_\infty}$ exhibits strong system-size effects~\cite{SupMat}, and decreases to zero with increasing size, as predicted by ETH for an effectively free spin. 
The phase diagram as extracted from this measurement is shown in Fig.~\ref{fig:phases}.
In the remainder, we  will work at a fixed disorder strength $W=3.0$, chosen sufficiently high that the system remains in the localized phase for all interaction strengths studied, with almost no finite-size effects ($L \gg \xi$, with $\xi$ the localization length).

\begin{figure}[t!]
\includegraphics[width=1\columnwidth]{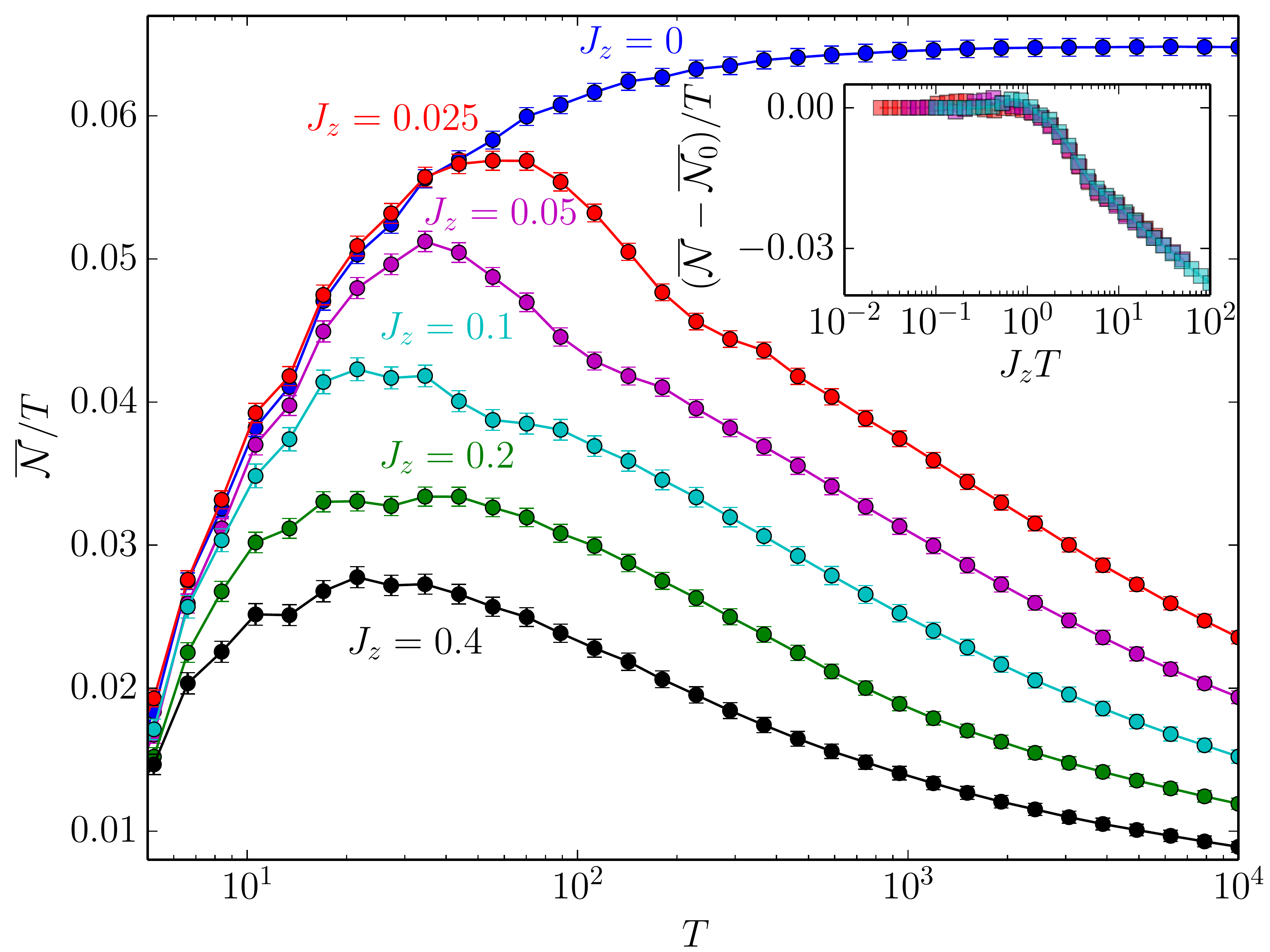}
\caption{\label{fig:revivals} {\bf Quantum Revivals.} Disorder-averaged revival rate $\overline{\mathcal{N}(T)}/T$ as function of total time, $T$.  Upon adding interactions of strength $J_z $, revivals are suppressed beyond $T^* \sim J_z ^{-1}$. (Inset) The same data collapses onto a universal curve when plotted against $J_z  T$, with ${\cal N}_0(T) = \left. {\cal N} (T) \right|_{J_z=0}$.}
\end{figure}

{\bf  Revivals.} 
After simulating the dynamics following the quench, we analyze the time series for the qubit magnetization $\langle S^z(t)\rangle$ (see Fig.~\ref{fig:Setup}(c)), identifying revivals in the magnetization~\cite{SupMat} (see also~\cite{PhysRevA.87.013424} and~\cite{PhysRevLett.100.100501,PhysRevLett.110.230601} for related proposals in the context of quantum phase transitions). Fig.~\ref{fig:revivals} shows the resulting number of revivals $\mathcal{N}(T)$ on the total time of evolution $T$ for an $L=10$ site chain for the specified disorder strength ($W=3.0$)  and for relatively strong interactions $J_z  \lesssim 0.4$.
 Clearly, the results depend sensitively on the presence of interactions. In the non-interacting, Anderson-localized phase, the revival rate $\mathcal{N}(T)/T$ grows with time until it reaches a constant value.  Upon adding interactions of strength $J_z $, the revival rate is strongly suppressed, beginning at a time $T^*\sim J_z ^{-1}$ (Fig.~\ref{fig:revivals}, and inset); this is also the time scale corresponding to onset of logarithmic entanglement growth previously reported. Finally, and most strikingly, the data collapses onto a universal curve when time is measured in units of $J_z^{-1}$.

We now demonstrate that the suppression of revivals traces its origin to the same dephasing mechanism that is responsible for entanglement growth. Following Refs.~\onlinecite{PhysRevLett.111.127201,PhysRevB.90.174202,2013arXiv1307.0507S, BauerNayak}, the MBL phase can be understood in terms of a set of conserved integrals of motion. (Quantum integrable systems that have an infinite set of conserved local integrals of motion also fail to obey ETH~\cite{Rigol:2008kq, PhysRevLett.98.050405} for sufficiently large subsystems, see {\it e.g.} Refs.~\onlinecite{PhysRevB.87.245107, PhysRevA.87.042105}, but arbitrarily weak generic perturbations restore ergodicity to these~\cite{Narozhny:1998}).
The $\tau^z$s are  exponentially localized in terms of their overlap with the original degrees of freedom $\sigma^\mu_{j}$ (physical bits or `p-bits'), and hence are termed localized bits (`l-bits'). Dephasing occurs solely due to the exponentially weak interactions between the spins, so that an effective Hamiltonian for the MBL system is \be\label{eq:MBLham}
H_\text{MBL} = \sum_i \omega_i \tau^z_i + \sum_{i,j}\mathcal{J}_{ij} \tau^z_{i}\tau^z_j +\ldots, 
\ee
where $\mathcal{J}_{ij}  \sim J_z  e^{-|i-j|/\xi}$ and the ellipsis denotes higher-order ($n\geq 3$-body) interactions. (Here and below, we take $\vec{\sigma}_0 \equiv \vec{S}$ for conciseness of the resulting expressions.)  Neglect of the higher-order terms is strictly justified deep in the MBL phase, but we find that the functional forms to be derived from (\ref{eq:MBLham}) apply numerically over nearly the whole phase.

 To study the suppression of revivals  it suffices to observe that the non-interacting revivals are governed only by frequencies $\omega_i$ of the $N\sim \xi$ 
l-bits  $\tau_i^z$ (see eq.~\eqref{eq:MBLham} with $J_z=0$) that have significant overlap with the qubit ($N$ will change depending on the specific choice of observable). Since the spectrum in the non-interacting limit is additive, it suffices to consider these $N$ levels, and require for instance~\cite{PhysRev.107.337}
$\sum_{i=1}^{N} |1-\cos(2\pi \omega_i t)| <\delta $ with $\delta$ a small parameter
in order that the many-body wavefunction experiences an (approximate) revival.
We will denote the corresponding disorder-averaged revival rate  $\Gamma_0(T,N)$ defined as the ratio of the number of such revivals in the time window $[0,T]$ to the total time $T$. This is clearly a decreasing function of $N$, 
but the dependence of $\Gamma_0(T,N)$ on the time $T$ is complicated and depends on the statistics of the $\omega_i$.
 If we now turn on interactions, then nearby orbitals experience random Hartree level shifts as a consequence of the $\mathcal{J}_{ij}$ term. The corresponding energy splitting of levels $\omega_i, \omega_j$ takes the form $\delta {\omega}_{ij} \sim J_z  e^{-|i-j|/\xi}$. For times $T\ll J_z ^{-1}$, the splitting is unimportant and does not significantly conflict with the revival criterion.
However, for $T\gtrsim J_z ^{-1}$, the Hartree shift of each nearest-neighbor pair is appreciable enough that, in effect, an additional  frequency  enters the revival criterion.  
 As $T$ increases further, each pair separated by distance $x$ leads to an additional frequency entering the revival criterion when $T\gtrsim e^{x/\xi}/J_z $, so that at time $t$ the appropriate revival rate is roughly $\Gamma_0(T, N+ \alpha\log J_z  T )$. 
 Thus, we find for the suppression of revivals relative to the non-interacting case $\frac{\overline{\mathcal{N} -\mathcal{N}_0}}{T} \approx \Gamma_0(T, N+ \alpha\log J_z  T) -  \Gamma_0(T, N)$.
This is not a universal function of $\log(J_z  T)$, due to the explicit dependence of $\Gamma_0$ on $T$. However, we argue that for strong disorder this dependence is only due to the randomness in the frequencies and as such is only weakly dependent on the {\it number} of independent frequencies, $N$. Therefore, we may write $\Gamma_0(T,N) \approx \gamma(T) + \nu(N)$ up to small corrections, so that
  \be
\frac{\overline{\mathcal{N} -\mathcal{N}_0}}{T} \approx \nu(N+ \alpha\log J_z  T) -  \nu(N)  ,
 \ee
which is a universal function of $\log J_z  T$, consistent with the collapse in Fig.~\ref{fig:revivals}. For $ \alpha \log J_z  T \sim \xi \log J_z T \ll 1$, we see that ${\overline{\mathcal{N} -\mathcal{N}_0}}/{T}  \approx - \alpha|\nu'(N)|\log J_z  T$. 

Clearly, aspects of the preceding analysis are non-universal  -- for instance, the precise value of $\mathcal{N}(T)$ will depend on the specific choice of time step $\Delta t$ and our algorithm for counting revivals. Note that this argument does not depend on precisely how the revival rate depends on the number of frequencies, although in the long-time limit one expects an exponential dependence. However, the mechanism behind the revival rate suppression traces its origin to the same hierarchical structure of the dynamics responsible for entanglement growth and leads to a similar logarithmic time dependence. Thus, the suppression of revivals by interactions is a universal signature in accord with the caricature of MBL systems as `localized but dephasing'~\cite{PhysRevB.90.174202}, and as such reveals the intrinsically interacting nature of the MBL phase. 
\begin{center}
\begin{table}
\begin{center}
  \begin{tabular}{| c| c | c |c | c|}
    \hline
  	{\bf Type} & Dephasing & Dissipation &$\,\,\,\,\,\,{N(t)}$ &${S_\text{EE}(t)}$\\
	\hline
	Anderson Ins. &  $\times$ & $\times$ &$\sim \xi$ & $\sim \xi$\\
	Many-body loc. & $\checkmark$ & $\times$ & $\sim\log J_z t $ &$\sim \log J_z t$\\
	Ergodic (ETH)  & $\checkmark$ & $\checkmark$ & $\,\,\sim t^\alpha$ &$\sim t$\\
      \hline
  \end{tabular}
  \end{center}
  \caption{\label{tab:Osc}{\bf Differences between MBL, Anderson-localized and ergodic phases} in terms of dephasing and dissipation, and their asymptotic  effective phase space volume $N(t)$ 
 governing revivals and growth of bipartite entanglement entropy $S_\text{EE}(t)$ in the limit $L \to \infty$. Deep in the MBL phase, the latter two quantities experience logarithmic growth attributable to dephasing. In the ergodic phase, $\alpha > 0$ could depend on the details of the system, {\it e.g.} whether it is ballistic or diffusive, while the entanglement grows linearly~\cite{PhysRevLett.111.127205}.}
  \end{table}
  \end{center}

{\bf Experiments.}  As we have already observed, ultracold atomic gases provide a natural experimental setting in which to explore the question of many-body localization, as they circumvent the problem, endemic to solid-state systems, of isolation from external sources of equilibration~\cite{RevModPhys.80.885,Kinoshita:2006,Yurovsky:2008,kondov2011three,schneider2012fermionic,gring2012relaxation,d2013quantum,PhysRevLett.113.217201}.  In addition, they possess a high degree of tunability: the strength of the interactions may be controlled by utilizing Feshbach resonances, and quenched disorder may be implemented by using a `speckle pattern' generated by a stationary configuration of laser intensity distributions~\cite{goodman2007speckle}.  It is also possible to selectively tune the fields at selected sites of an optical lattice, enabling the identification of one or more sites as the `qubit' in our analysis.  It should be noted that solid-state systems are starting to achieve a similar level of tunability at least in small systems: up to five interacting ``transmon qubits'' can now be manipulated with high fidelity~\cite{martinis}.

In order to study quantum revivals in either setting, we must measure the state of the spin at the selected site at time $t$, which is an inherently destructive measurement. Therefore, many repetitions of the experiment will have to be performed with a single realization of disorder. As the speckle pattern can be reproduced and changed on-demand, the necessary repetition does not pose a fundamental obstacle beyond the additional time required to make multiple measurements. It is worth also noting that we expect the revival pattern to persist even if the initial state of the system is not exactly the same between repetitions. We remark that the distinct behavior of quantum revivals in ergodic, MBL, and Anderson-localized systems persists  even for a single typical disorder configuration -- see Fig.~\ref{fig:Setup}(b).

{\bf Discussion.}  In this paper, we have connected a fundamental feature of many-body dynamics in a finite phase space --- namely, the quantum `Poincare' recurrence probability --- to the question of thermalization in isolated systems. We have shown that revivals of a single qubit weakly coupled to a disordered ``reservoir'' allows one to distinguish between Anderson-localized, many-body localized, and ergodic phases of the reservoir. The interaction-induced dephasing characteristic of MBL systems is responsible for the distinction in the `effective phase space volume' for dynamics in the two different localized phases: in the MBL case, it leads to a logarithmic growth in time of the effective number of frequencies that must synchronize in order for the qubit to revive. In the ergodic phase, the revival probability is exponentially small as $L\rightarrow\infty$ on the time scales studied, and thus $\overline{\mathcal{N}(T)}$ is nearly vanishing. Table \ref{tab:Osc} summarizes the  distinctions between ergodic (satisfying ETH), MBL and Anderson localized phases in terms of their revival dynamics and entanglement growth at long times. 

We were led to consider revivals initially because they were found to be a more sensitive probe, both of localization as well as the nature of the localized phase, as compared to measures such as the power spectrum of local observables ({\it i.e.} the Fourier transform of the time series of $\langle \mathcal{O}(t)\rangle$) or other standard quantities~\cite{PhysRevB.90.064203}.
 As an added bonus, the revival probability is simple to define and depends only on the magnetization which is straightforward to measure. An appealing feature of using such dynamical probes is that they allow a single measurement to establish both the nature of dephasing -- via revival analysis -- and dissipation, encoded in the long-time steady-state average magnetization. Given the paucity and technical complications of existing probes of ergodicity breaking in general and many-body localization in particular, we expect that these features make revival analysis an appealing route to establishing the existence of the MBL phase in real systems.

{\bf Acknowledgements.}  We thank E.~Altman, B.~Bauer, E.~Demler, V.~Oganesyan, A.C.~Potter,  R.~Vosk, M.~Zaletel  and especially J.~Bardarson,  S.~Gopalakrishnan,  and R.~Nandkishore  for insightful discussions and comments on the manuscript, and Mandy Muller for assistance preparing the figures. We acknowledge support from the Simons Foundation (S.A.P. and J.E.M.), UC Irvine startup funds (S.A.P.), the
Quantum Materials program of LBNL (R.V.) and NSF grant DMR-1206515 (J.E.M.).

\bibliography{MBL}

\end{document}


\title{Supplemental Material for ``Quantum revivals and many-body localization''}
\author {R. Vasseur}
\affiliation{Department of Physics, University of California, Berkeley, CA 94720, USA}
\affiliation{Materials Science Division, Lawrence Berkeley National Laboratories, Berkeley, CA 94720}
\author{S. A. Parameswaran}
\affiliation{Department of Physics and Astronomy, University of California, Irvine, CA 92697, USA}
\author{J. E. Moore}
\affiliation{Department of Physics, University of California, Berkeley, CA 94720, USA}
\affiliation{Materials Science Division, Lawrence Berkeley National Laboratories, Berkeley, CA 94720}

\date{\today}

\maketitle
\renewcommand{\thesection}{\text{S}\Roman{section}}
\renewcommand\thefigure{S.\arabic{figure}}    

\onecolumngrid



 \section{\label{methods:numtech}Numerical Techniques} In our analysis of revivals, we employ  time-evolving block decimation (TEBD)\cite{PhysRevLett.91.147902}, an algorithm closely related to the density-matrix renormalization group\cite{PhysRevLett.69.2863,Schollwock201196}  that uses the matrix-product state (MPS) representation of wavefunctions that is particularly well-adapted to the slow (logarithmic) growth of entanglement in the MBL phase. We perform a fourth-order order Suzuki-Trotter decomposition of the short-time propagator $U(\Delta t) = \exp(-i\Delta t H)$ with a time step $\Delta  t= 0.1$, and control the growth of bond dimension in successive time steps by truncating low-weight states in a Schmidt decomposition. The TEBD algorithm we use here requires that the neglected weight be $<10^{-7}$; in practice, this means that for sufficiently small systems and at sufficiently long times, the MPS bond dimension saturates so that TEBD becomes exact -- up to the Trotter error -- and has no efficiency gain over exact diagonalization (ED). However, for short to intermediate times TEBD is extremely efficient, and for these time scales, allows us to extend part of our analysis to system sizes beyond the reach of ED. In particular, we are able to obtain long- (but not infinite-) time behavior of the magnetization for $L=40$.
We average over a large number (typically  $10^3-10^4$) instances of disorder to obtain reliable statistics and to reduce sampling error. 

\section{\label{supmat}Finite-Size Scaling Analysis}
We briefly summarize how we obtain the phase diagram in Fig.~2 from finite-size scaling of the magnetization $S_z(t)$ of the qubit. While in some cases, the difference between ergodic and many-body localized phases is sufficiently sharp that it may be asserted from times series as in Fig.~1(c), a more thorough analysis is provided by the finite-size scaling of the infinite-time value  $S_z^\infty$ of the qubit magnetization, computed from eq.~(3) using exact diagonalization. We have verified that the values of $S_z^\infty$ obtained in this manner were consistent with the curves of $S_z(t)$ obtained from TEBD. 

In the ergodic phase, we expect $S_z^\infty$ to be zero, as appropriate for a free spin that is effectively at infinite temperature. However, in practice $S_z^\infty$ is always finite for finite systems -- barring  some pathological cases with $W=0$ in the fixed $S_z=0$ sector, attributable to additional symmetries present in this limit -- and a detailed finite-size scaling analysis is required. Deep in the MBL phase, we find that $S_z^\infty$ does not depend on system size (up to error bars), clearly indicating of a localized system with localization length smaller than the total system size. Closer to the transition, we find weak finite-size scaling effects that can be fitted very well by $1/L$ corrections. In particular, note that  $S_z^\infty(L)$ is a {\it convex} function of $L$, that is, the derivative $ {\rm d} S_z^\infty/ {\rm d} L <0 $ is an increasing function of $L$, which goes to zero as $L\to \infty$. We attribute this behavior to the MBL phase. On the other hand, for weak disorder and/or strong interactions, we find that $S_z^\infty(L)$ shows very strong finite size effects, which cannot be fitted by $1/L$ corrections only, and which start as a {\it concave} function of $L$: the derivative $ {\rm d} S_z^\infty/ {\rm d} L <0 $ is a decreasing function of $L$. We conjecture that the function $S_z^\infty(L)$ has an inflection point for some characteristic value of $L$, necessary in order that $S_z^\infty(L)$ have a physically valid limit as $L \to \infty$. In the ergodic phase, we expect this limit to be $0$, although this cannot be assessed from our numerics alone. We leave the analysis of this potentially interesting crossover as a function of $L$ for future work; for now, we simply assert that  this scaling behavior should be associated with an ergodic phase. 
\begin{figure*}[t] 
 \centering
 \includegraphics[width=0.49\textwidth]{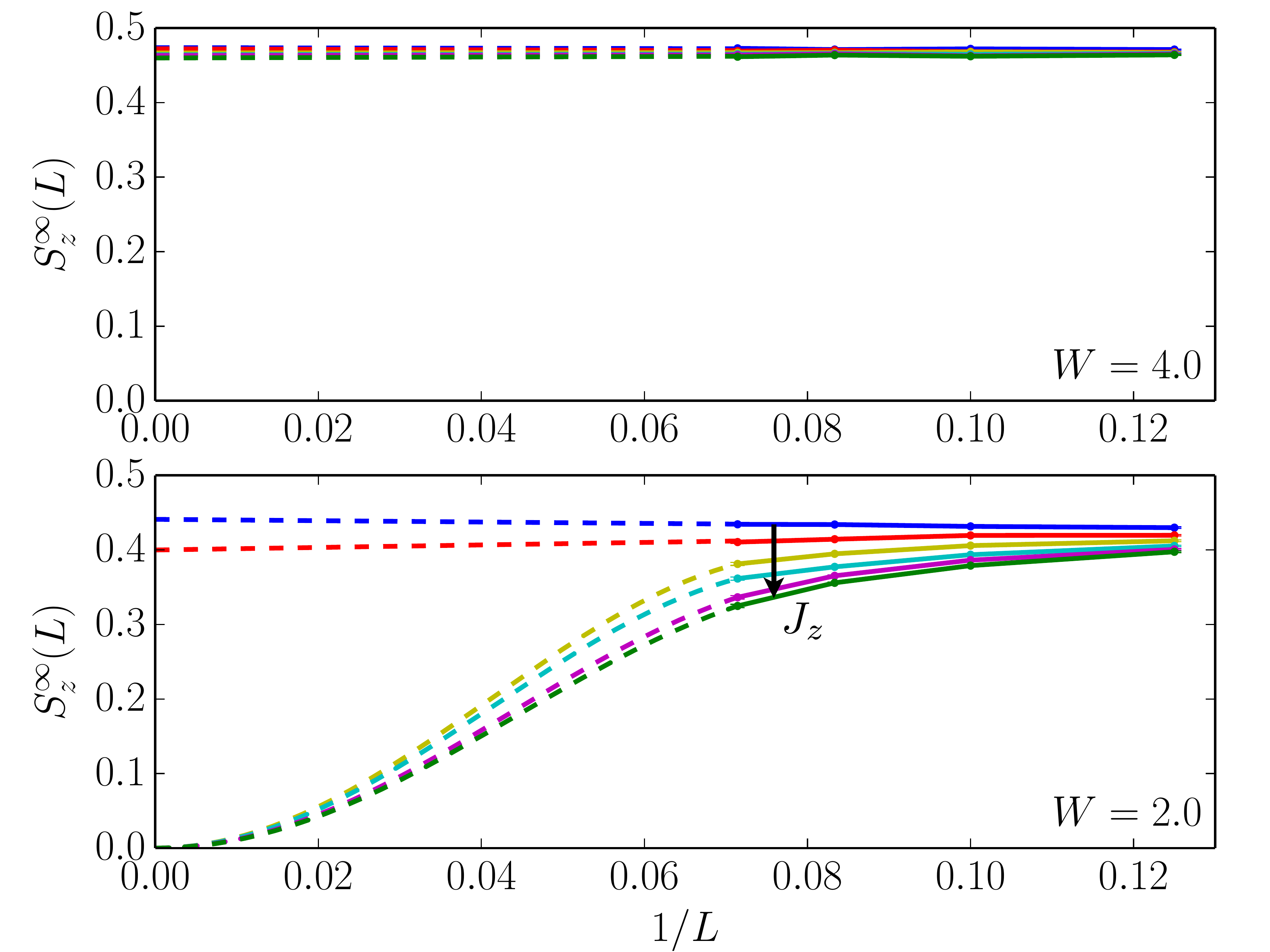}
 \includegraphics[width=0.49\textwidth]{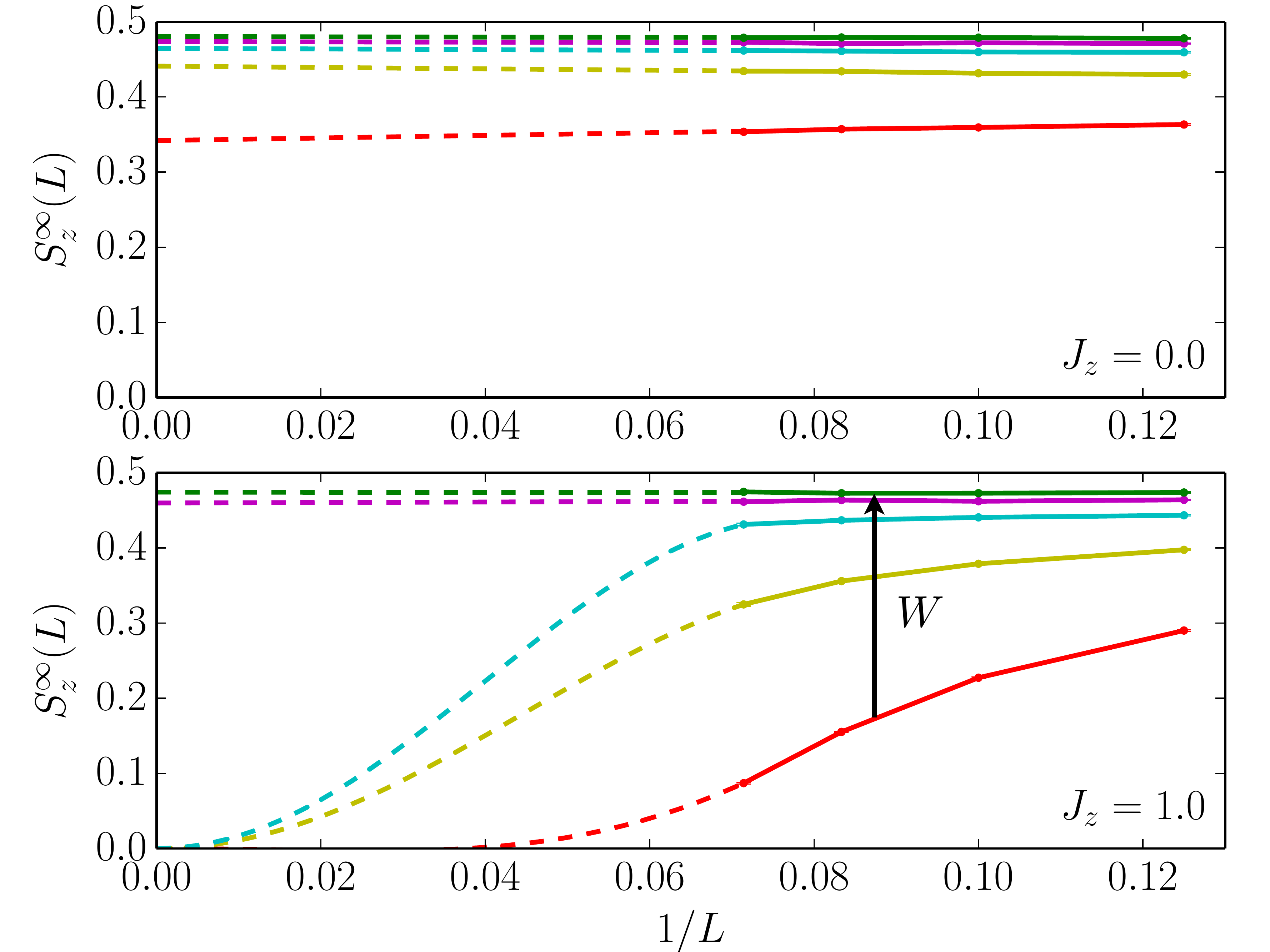}
 \caption{Infinite-time value $S_z^\infty$ of the magnetization of the qubit, as a function of the total system size $L$.}
  \label{fig:FSS}
\end{figure*}

This criterion on the nature of finite-size scaling of $S_z^\infty(L) $ serves as a clear operational definition of the MBL and ergodic phases. While it does not allow us to probe very precisely where the transition lies -- some points in the phase diagram cannot be associated with either phase owing to error bars -- it is sufficient to obtain the general form of the phase diagram shown in Fig.~2. Further examples of finite size scaling are shown in Fig.~\ref{fig:FSS}.

We have verified that the resulting phase diagram does not depend on the coupling $\lambda$ between the qubit and the chain. In particular, our method also applies if $\lambda = J_\perp=1$ so that the qubit is essentially an ordinary spin, albeit one not subject to a local magnetic field. We have also verified that our predictions do not depend on the precise nature of the initial state, as long as it corresponds to a very high temperature: the same phase diagram was obtained starting from a N\'eel state,  a random product state that is an eigenstate of arbitrary total magnetization, or a random product state of zero total magnetization.

\section{\label{methods:rev} Revival Counting} 
We obtain $\mathcal{N}(T)$ for each disorder realization as follows: from the time series of  $\langle S^z(t)\rangle$ obtained via TEBD for $0<t<T$, we obtain a time-ordered sequence of local maxima, $\{S^z_{\max}(t_0\equiv 0)\equiv \frac12,S^z_{\max}(t_1), \ldots\}$; every revival will correspond to one of these, but not all of these will be true revivals. Given that a revival occurs at $t=t_j$, we may obtain the next revival by examining successive later maxima at $t_{j+1}, t_{j+2},\ldots$ and for each of these computing the quantity \be\chi \equiv \frac{S^z_{\max}(t_0) - S^z_{\max}(t_k) }{S^z_{\max}(t_0) - \min\{\langle S^z(t)\rangle | t_j< t  < t_k\}}\ee
and accepting $t_k$ as a revival  if $\chi <\epsilon$, or else proceeding to $t_{k+1}$. Here, $\epsilon =0.05$ is chosen sufficiently small to exclude instances where the magnetization partially recovers without 
first deviating substantially from its initial value. The above algorithm may be initialized by taking $t_j =t_0$ in the first step, and subsequently proceeds iteratively. Finally, $\overline{\mathcal{N}(T)}$ is obtained by disorder averaging.

\bibliography{MBL}